\documentclass[twocolumn,prb,amsmath,amssymb,amsfonts,superscriptaddress,floatfix]{revtex4}
\usepackage{graphicx}
\usepackage{dcolumn}
\usepackage{bm}
\usepackage{rotating}
\usepackage{color}
\usepackage{amsmath}
\usepackage{comment}

\def\be{\begin{equation}}
\def\ee{\end{equation}}
\def\bd{\begin{displaymath}}
\def\ed{\end{displaymath}}
\def\-{\phantom{-}}
\usepackage{lipsum}

\begin{document}


\title{Designing a path towards superconductivity through magnetic exchange in transition-metal intercalated bilayer graphene}

\author{Kevin P. Lucht}
\affiliation{Condensed Matter Theory - National High Magnetic Field Laboratory (NHMFL), Florida State University, Tallahassee, FL, 32310, USA}
\affiliation{Department of Scientific Computing, Materials Science and Engineering, High Performance Materials Institute (HPMI), Florida State University, Tallahassee, Florida, 32310, USA.}
\affiliation{Department of Physics, College of Arts and Science,Florida State University, Tallahassee FL, 32310, USA.}

\author{A. D. Mahabir}
\affiliation{Department of Physics, University of North Florida, Jacksonville, FL 32224, USA}

\author{A. Alcantara}
\affiliation{Department of Physics, University of North Florida, Jacksonville, FL 32224, USA}

\author{A. V. Balatsky}
\affiliation
{Nordic Institute for Theoretical Physics, KTH Royal Institute of Technology and Stockholm University, Roslagstullsbacken 23, 106 91 Stockholm, Sweden}

\author{Jose L. Mendoza-Cortes}
\email{mendoza@magnet.fsu.edu}
\affiliation{Condensed Matter Theory - National High Magnetic Field Laboratory (NHMFL), Florida State University, Tallahassee, FL, 32310, USA}
\affiliation{Department of Scientific Computing, Materials Science and Engineering, High Performance Materials Institute (HPMI), Florida State University, Tallahassee, Florida, 32310, USA.}
\affiliation{Department of Physics, College of Arts and Science,Florida State University, Tallahassee FL, 32310, USA.}

\author{J. T. Haraldsen}
\email{j.t.haraldsen@unf.edu}
\affiliation{Department of Physics, University of North Florida, Jacksonville, FL 32224, USA}

\date{\today}

\begin{abstract}

This study examines the potential of superconductivity in transition metal (TM) intercalated bilayer graphene through a systematic study of the electronic and magnetic properties. We determine the electronic structure for all first row TM elements in the stable honeycomb configuration between two layers of graphene using density functional theory (DFT). Through an analysis of the electron density, we assess the induction of the magnetic moment in each case, where a comparison of the ferromagnetic and antiferromagnetic configurations allow us to ascertain an estimated exchange coupling between the transition-metal elements. 
By analyzing the electronic properties, we find that the carbon $p$-bands are degenerate with the TM $d$-bands and form an electron pocket below the Fermi energy at the $\Gamma-$point. These hybridized bands are analogous to the carbon $p$-band effect that produces superconductivity in intercalated graphite with alkali and alkaline-earth metals. Furthermore, since the bands are hybridized with the TM $d$-bands, their magnetic properties may provide bosonic modes from their spin-coupling to preserve the unique linear dispersion present in monolayer graphene. This study provides a designing route by using TMs for tuning magneto-electric Dirac materials and will encourage future experimental studies to further the fundamental knowledge of unconventional superconductivity.

\end{abstract}

\maketitle


\section{Introduction}

Graphene is one of the most studied materials in condensed matter physics due to the possible technological and industrial applications ranging from nanopore and spintronic devices to transistors and semiconducting technologies.\cite{Trau:07,Schw:10,Tomb:07,Schn:10,Bane:13,han:14} Beyond its myriad of applications, the recent discovery of superconductivity in twisted bilayer graphene through magic angles has sparked a theoretical challenge to understand the potential superconducting mechanism.\cite{Cao:18} With graphene's wide range of interesting phenomena and applications, it has maintained its spotlight in research for over a decade.

Due to its electronic structure, graphene is known as a Dirac material, where a single honeycomb lattice of carbon atoms has an inversion symmetry from its two sublattice structure. This symmetry produces the formation of a fermionic linear dispersion in the electronic structure at two points of high symmetry known as the Dirac cone.\cite{wehl:14} While the Dirac cone is a fairly robust electronic state, it can be destroyed through chemical substitution, spin-orbit coupling, or even the presence of another graphene layer.\cite{Yao:07,wehl:14,krac:11,chan:11}

\begin{figure}
\includegraphics[width=0.5 \linewidth]{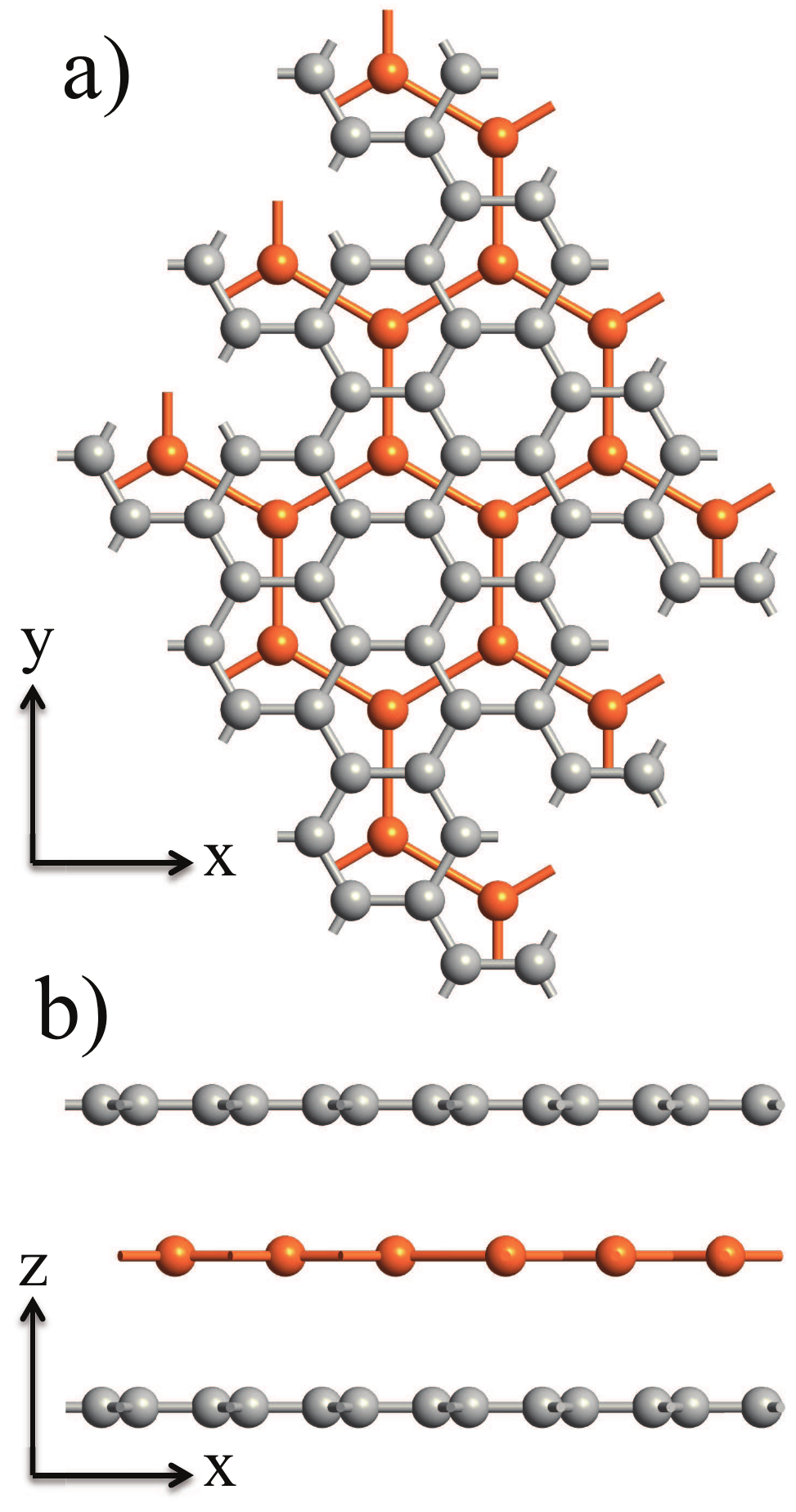}
\caption{Illustrations of the (a) $xy$-plane and (b) $xz$-plane structures of transition metal (TM) filled graphene sandwiches. Coloring Scheme: C, grey; TM, orange.}
\label{structure}
\end{figure}

While all the interest in graphene is focused on its electronic properties, the fragility of graphene's intrinsic properties makes it difficult to be used in device applications. These issues have led to the exploration into possible alternatives, notably low dimensional boron compounds.\cite{Ma:16,Oganov:09} However, two-dimensional compounds exhibiting graphene's electronic properties are more challenging to synthesize. Alternatively, stacked graphene structures provide additional properties such as recently explored quantum hall phase transitions which make them suitable for broader tunability.\cite{Zibrov:18,Li:17,Jacak:17} With regards to Dirac physics, stacked graphene offers no benefit as the energetically favorable AB stacking breaks the inversion symmetry and effectively eliminates the degeneracy at the band crossing. This effect is apparent through the formation of a band gap as the electronic dispersion relation becomes parabolic rather than linear.\cite{ohta:11} However, bilayer graphene is a highly tunable material for which the bandgap can be tuned by introducing a chemical dopant, by applying a gate voltage, or by maintaining a difference in electronic potential.\cite{Yu:09,Coletti:10,Castro:07}

The effects of metal adsorption onto graphene layers has been an ongoing area of study due to the properties that result from their interaction with graphene, such as the quantum anomalous Hall effect and emerging Hall states from symmetry breaking. \cite{ding:11,Zhang:2012} In general, it is well documented that transition metals elicit magnetic properties and lower the Fermi energy of graphene due to the presence of $d$-orbital electrons.\cite{zhou:11,yuchen:11,mao:11,sutter:11}
Recently, there has been a lot of interest in magnetic Dirac modes, where nearest neighbor interactions of free electron spins can couple to form bosonic models. These modes contain a Dirac cone that is visible in the bosonic spectrum with converging modes that can have similar symmetries to its fermionic partner.\cite{Fransson:16,boyk:18} Therefore, an important model is one where fermionic and bosonic Dirac symmetries can co-exist.
For many years, there has been a significant push to have magnetic coupling in graphene layered structures through surface adsorption or inner layer intercalation. This is most evident by the research on yttrium iron garnet (YIG)/graphene heterostructures and even magnetic edge states in disordered graphene systems.\cite{Wang:2015,Hallal:2017} However, to design this kind of multifunctional Dirac material, one has to consider materials that provide compatible geometric symmetries. In the study by Boyko \textit{et al.}\cite{boyk:18}, Dirac bosons can be observed through magnetic interaction in a honeycomb lattice. Therefore, in this study, we examine transition-metals intercalated into bilayer graphene to investigate the potential of bosonic modes. 

Furthermore, superconductivity has been a heavily discussed possibility in metal-intercalated graphene, which has gained considerable experimental support in recent years with the number of published superconducting structures continuing to grow. The evidence for metal-intercalated graphite was provided for Ca and Yb \cite{Weller:05} with theoretical insight quick to follow. From first-principles studies, \cite{Csanyi:05,mazin:11} it was conjectured that superconductivity produced from Ca and Yb intercalation was induced from interactions from the metal $p$-orbitals and graphite sheet $\pi^*$ orbitals which were caused by carbon $p$-band crossing at the Fermi energy. Additionally, Calandra $et~al.$ \cite{Calandra:05} utilized first-principles methods to suggest that superconductivity was a phonon-mediated process. Later, experimental evidence was presented for superconductivity in Li, Ca, and K intercalated in few-layer graphene (FLG) with phonon-mediation identified as the key process in Ca-intercalated graphene. \cite{Margine:2016} On the computational side, Sr and Ba have been suggested as potential superconducting intercalates while also reaffirming Li and Ca superconductivity. \cite{Kaneko:17,Lee:12} An interesting result from first-principle calculations by Kaneko \textit{et al.}\cite{Kaneko:17} is that, similar to the $p$-band crossing in Yb and Ca intercalation in graphite; Ca, Sr, and Ba superconductors have a $d$-band crossing at the Fermi energy. This result suggests that $d$-band crossing could be a reliable indicator of superconductivity in graphene.

Thus, few-layer graphene with metal intercalation has yielded a diverse array of applications and unique physical properties, which has continued to expand in recent years. Most notably is superconducting intercalated graphene and from the recent discovery of magic angles in bilayer and twisted graphene.\cite{Cao:18} However, studies in intercalated graphene have mainly focused on alkali and alkaline metals although a possible route to their superconductivity is from $d$-band crossing.\cite{Kaneko:17} This prompt us to hypothesize that transition-metal elements may also serve as potential candidates for facilitating superconductivity in bilayer graphene. A further motivator for transition-metal elements stems from their magnetic properties, which could provide bosonic modes from their spin-coupling while also serving as a structural brace to preserve the unique linear dispersion present in monolayer graphene. With both the electronic and magnetic properties, intercalatation of transition-metals could be suitable for magneto-electric Dirac materials. 

In this study, we use density functional theory (DFT) to study the electronic and magnetic behavior of bilayer graphene intercalated with all the 3$d$ transition-metal elements. In the analysis of the electronic structure, we find evidence that the intercalation of these metals may also lead to superconducting states, similar to Ca, K, and Li intercalated systems. Furthermore, we find that only two of the ten transition-metal elements provide viable candidates for magneto-electric Dirac materials with the formation of ferromagnetic states.

\begin{figure}
\includegraphics[width=1.0 \linewidth]{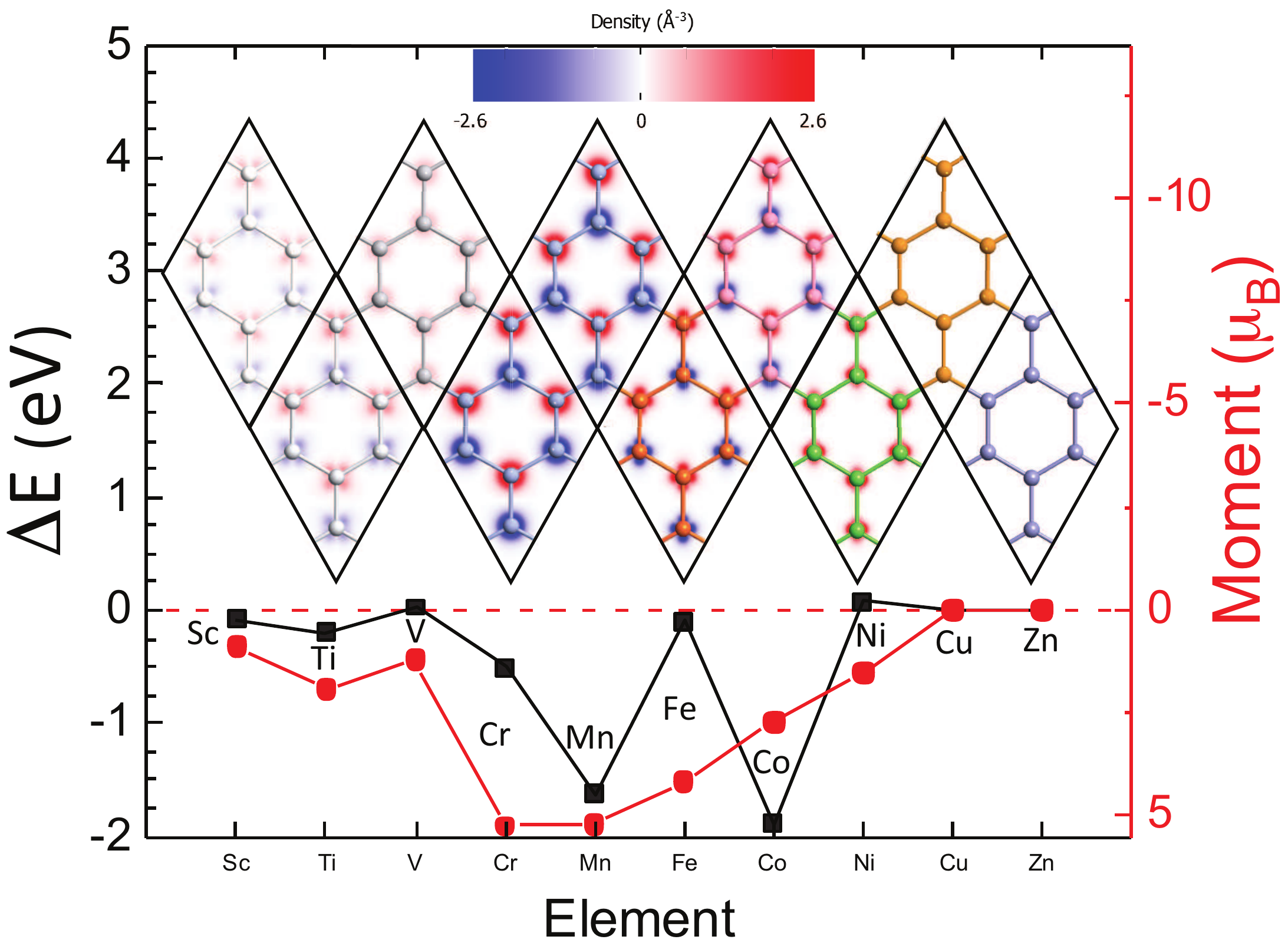}
\caption{Energy difference (left axis) and magnetic moment (right axis) as a function of transition-metal element. The top inset shows the electron density for each transition-metal layer.}
\label{ED-MM-DE}
\end{figure}

\section{Computational Methodology}

Starting with a unit cell consisting of twelve carbon atoms and two transition metal atoms; a honeycomb lattice of transition metal atoms was sandwiched between two graphene layers (Figure \ref{structure}). Recent numerical results show that AA-stacked graphene sheets with transition metal atom intercalated and centered within a honeycomb are the most energetically favorable.\cite{Sri:18}  A computational analysis was performed using Density Functional Theory (DFT) provided by Atomistix Toolkit\cite{quantumwise}, where the structures were fully optimized using Limited-Memory Broyden-Fletcher-Goldfarb-Shanno methods\cite{sole:02}. Additionally, the DFT calculations allowed spin polarizability by using a Spin-Polarized Generalized Gradient Approximation with the Perdew, Burke, and Ernzerhof (PBE) functional,\cite{perd:96} where a 30$\times$30$\times$1 $k$-point and tolerance of 10$^{-5}$ Hartrees were used. To examine the electronic and magnetic properties, we determined the electronic band structure, density of states, electron density, Mulliken population analysis, and total energy for the antiferromagnetic (AFM) and ferromagnetic (FM) configurations for all first row transition-metal elements.

\section{Magnetic Properties and Ground States}

The intercalation of transition-metal elements in between graphene layers provides a unique possibility of coupling between Dirac fermions and bosons. Given that the graphene layer provides a fermionic Dirac cone, the intercalation of a honeycomb structure with transition-metal elements allows for the possible formation of Dirac bosons, assuming the magnetic ground state is in the ferromagnetic configuration. Therefore, to analyze the magnetic structure, we model the spin states using a standard Heisenberg Hamiltonian without the presence of an external magnetic field,

\be
\mathcal{H} = -\frac{1}{2}\sum_{\substack{i,j \\ i \neq j}} J_{i,j}\Vec{S}_i \cdot \Vec{S}_j.
\ee

Here, $J_{i,j}$ is the exchange coupling between $S_i$ and $S_j$. Using a nearest-neighbor approximation such that $J_{i,j} = J$ with a Holstein-Primokoff expansion in $1/S$, we can find the classical energies for the ferromagnetic (FM) and antiferromagnetic (AFM) honeycomb lattice.\cite{boyk:18} Within a nearest-neighbor approximation, the classical energy difference ($\Delta E$) between the AFM and FM honeycomb lattice is $JS^2$, where $J$ is the exchange interaction, and $S$ is the classical spin.\cite{boyk:18} Using $\Delta E$, we can determine the magnetic ground state for each configuration since if $\Delta$E is positive, then the ground state is FM and if it is negative, then it is AFM.

\begin{figure}
\includegraphics[width=1.0\linewidth]{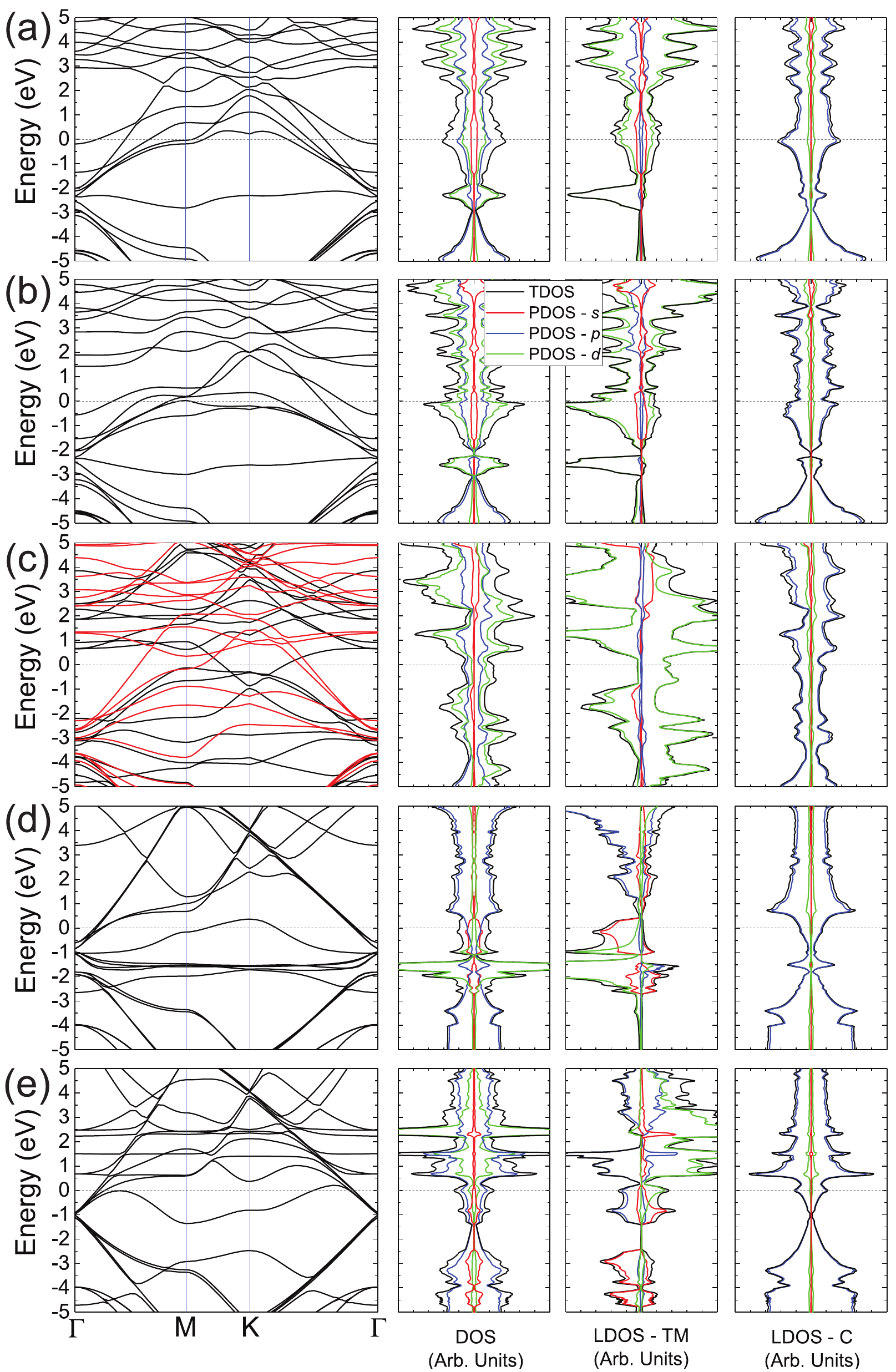}
\caption{Electronic band structure and the total, partial, and local density of states for (a) Sc, (b) Ti, (c) V, (d) Cr, and (e) Mn. All systems are in an antiferromagnetic ground state with exception for vanadium, which is ferromagnetic}
\label{Sc-Mn-BS-DOS}
\end{figure}

\begin{figure}
\includegraphics[width=1.0\linewidth]{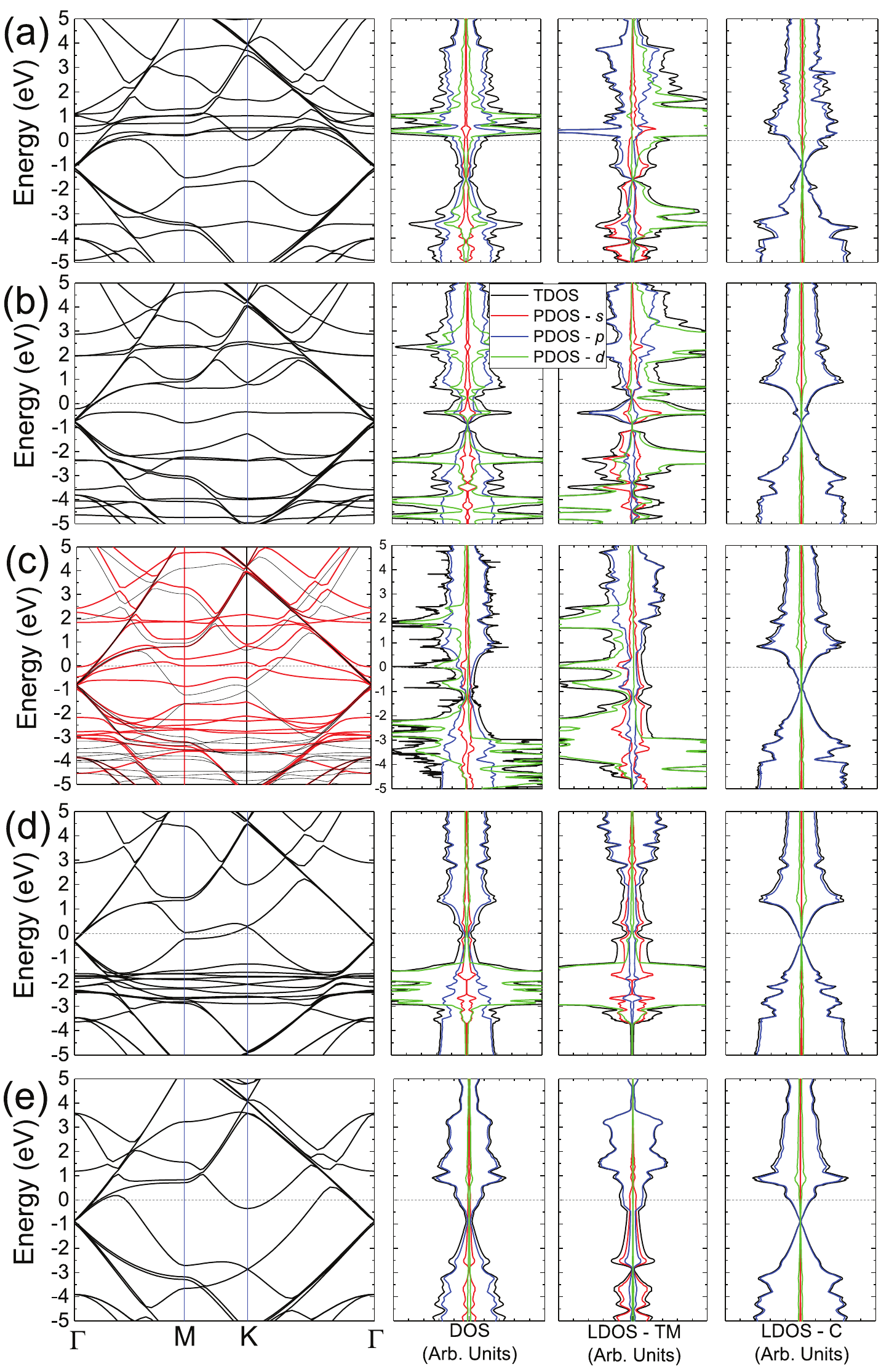}
\caption{Electronic band structure and the total, partial, and local density of states for (a) Fe, (b) Co, (c) Ni, (d) Cu, and (e) Zn. Fe and Co produce antiferromagnetic ground states, Ni is ferromagnetic, and Cu and Zn do not display any magnetic moment.}
\label{Fe-Zn-BS-DOS}
\end{figure}

Figure \ref{ED-MM-DE} shows the change in energy and calculated magnetic moment for all ten transition-metal graphene bilayer lattices. Furthermore, the inset shows the electron density difference, which details the ground state magnetic configurations. From the energy difference ($\Delta E$) between AFM and FM configurations, the ground state and estimated exchange interaction for all ten elements can be determined and compared. Here, it is clear that the middle transition metal elements (Cr, Mn, and Co) have a relatively large exchange interaction, while the other elements are either small (in meV range) or zero (Cu and Zn). Here, only V and Ni provide ferromagnetic ground states, which is necessary for the creation of Dirac bosons. 

The right side axis (red) of Figure \ref{ED-MM-DE} shows the calculated magnetic moment for each system. Sc, Ti, and V have magnetic moments close in value but the magnetic moment dramatically increases to 5 $\mu_B$ for Cr and Mn indicating a large number of unpaired electrons. From Mn to Ni, the magnetic moment steadily diminishes to reach zero for Cu and Zn, which is distinctly correlated to the electron density difference shown as blue (spin down) and red (spin up) in the structures above the curves. From the electron density difference, it is clear that the majority of elements have antiferromagnetic ground state configurations, while V and Ni are ferromagnetic. As the number of electron density increases among the elements, the magnetic moment changes depending on the number of unpaired electrons. However, the $d$ orbital electrons of Cu appear to have become delocalized in such a way that there is no resulting magnetic moment.

\section{Electronic Structure and Density of States}

Figures \ref{Sc-Mn-BS-DOS} and \ref{Fe-Zn-BS-DOS} show the electronic band structure and density of states for each transition-metal graphene bilayer system. Here, the density of states (DOS) includes the total DOS (TDOS), each orbital occupancy DOS ($s$-, $p$-, and $d$-PDOS), and local DOS (LDOS) for the TM and carbon atoms. Here, the spin down channel is to the left and spin up channel is on the right. The presence of fermionic and bosonic Dirac nodes seems to only occur in the Ni system, and therefore there is no valid way to assess the likelihood of fermionic/bosonic coupling. One distinct property of the presence of TM atoms in between the graphene sheets is a shift of the Dirac cone from the carbon below the Fermi level, which is independent of the TMs. In the Sc, Ti, V, and Cr cases, the electronic Dirac cone loses its linearity and becomes quadratic and even gapped indicating a strong chemical interaction between the graphene layers and the TM interlayer. However, in the other TM cases, the electronic Dirac cones begin to slowly reappear along the TM row of the periodic table until its emergence in Cu and Zn near the Fermi energy. 

A substantial shift in the electronic structure is apparent from Fig. \ref{Sc-Mn-BS-DOS}, where a sudden increase in electron density as well as in Fig. \ref{Fe-Zn-BS-DOS}  by the partial reformation of linear dispersion in the graphene layers between V and Cr. Before this point, it appears that TM $d$ orbital and carbon $p$ orbitals hybridize near the Dirac point which drastically shifts the Dirac cone down several eV. For V, the hybridization is strong enough to disrupt the linear dispersion in the graphene sheets. As discussed recently by Han et al.,\cite{Han:18} transition metal intercalation interacts with the carbons by forming $d-sp^2$ hybridize$d$ orbitals in addition to the breaking of $d$ orbital degeneracy in order to lower the occupational energy. Previous work on metals surfaces has also shown similar behavior that when graphene is chemisorbed onto a metal surface, where orbital hybridization breaks the Dirac cone while shifting it below the Fermi energy.\cite{Khom:09} The strength of the hybridization seems to impact the degree which the $d$ orbitals descend below the Fermi energy; with V, in particular, removes all resemblance of a Dirac cone in the graphene layer while supporting a high density of states from its $d$ orbitals. Consequently, its $d$ band crossing at the Fermi energy begins nearly 2 eV below the other TMs which is approximately measured in Figure \ref{A-B}b by the difference in energy between the A band at the $\Gamma-$point and the Fermi energy (set as the reference to zero).

Continuing along the $3d$ row of transition metals, the hybridization abruptly stops at Cr which is noted by the substantial electron density and magnetic moment shown in Figure \ref{ED-MM-DE}. From this point on, the interaction between the TMs and the graphene layers is mainly physical, designated by the sudden jump in c-axis distance shown in Figure \ref{Elattice} and the Dirac cone begins to reform. The decrease in the magnetic moment in Figure \ref{ED-MM-DE} is merely expressing that the electrons start to pair together until no magnetic moment is present. Partial hybridization still seems to be present in Cr, Mn, Fe, Co, and Ni as the Dirac cone has not completely reformed and remains around 1 eV below the Fermi energy. With little interaction between the TM $d$ orbitals and $p$ orbitals of graphene, the Dirac cone shifts to be within 0.5 eV of the Fermi energy for Cu. Unsurprisingly, Zn offers no hybridization with its fully occupied d orbitals, yet the physical interaction decreases the Dirac cone to be around 1 eV below the Fermi energy. Still, the presence and proximity of the Dirac cone to the Fermi energy in Cu and Zn suggests that they could be suitable Dirac materials.

Overall, the introduction of TM atoms into the bilayer structure of graphene seems to produce a shift in the fermionic Dirac cone to lower energy in the band structure and density of states. The TM atoms with few $d$ orbital electrons seem to hybridize with the bilayer graphene with results in quadratic band formation of the Dirac cone and lowering of band energies by several eV. However, beyond V, the TM atoms seem to suddenly prefer to only physically interact with the graphene layers which eventually restores the standard fermionic Dirac nature. Although this physical interaction is not as strong as the chemical interaction, it is still strong enough to lower the electronic band energies so that the Dirac cone is below the Fermi energy.

\section{Possibility for Intrinsic Superconductivity}

Superconductivity has been an ongoing exploration in graphene and graphene-based materials and has recently been invigorated due to the discovery of unconventional superconductivity due to magic angles in twisted bilayer graphene.\cite{Cao:18} Prior to the discovery of graphene, superconductive behavior was shown to exist in intercalated graphite, which exhibits superconductivity in the cases of Ca and Yb.\cite{Weller:05} This has also been observed in bilayered graphene for intercalated Li, K, and Ca. For graphite intercalation, the determination for superconductivity came from the conjecture that the chemical potential of graphene is shifted through the presence of the intercalated atoms.\cite{Csanyi:05} Most notable is a $p$ band crossing at the Fermi level and production of electron pockets at the $\Gamma$-point produced from lowering of the charge transfer between the metal and neighboring carbons. The mechanism for this behavior was shown to be controlled through layer separation and the occupancy of orbits.\cite{Csanyi:05} Similar results arise from the recent investigations of graphene intercalated systems with alkali and alkaline metals, as well as electronegative atoms such as N and O. \cite{Gong:10} 

\begin{figure}
\includegraphics[width=1.00\linewidth]{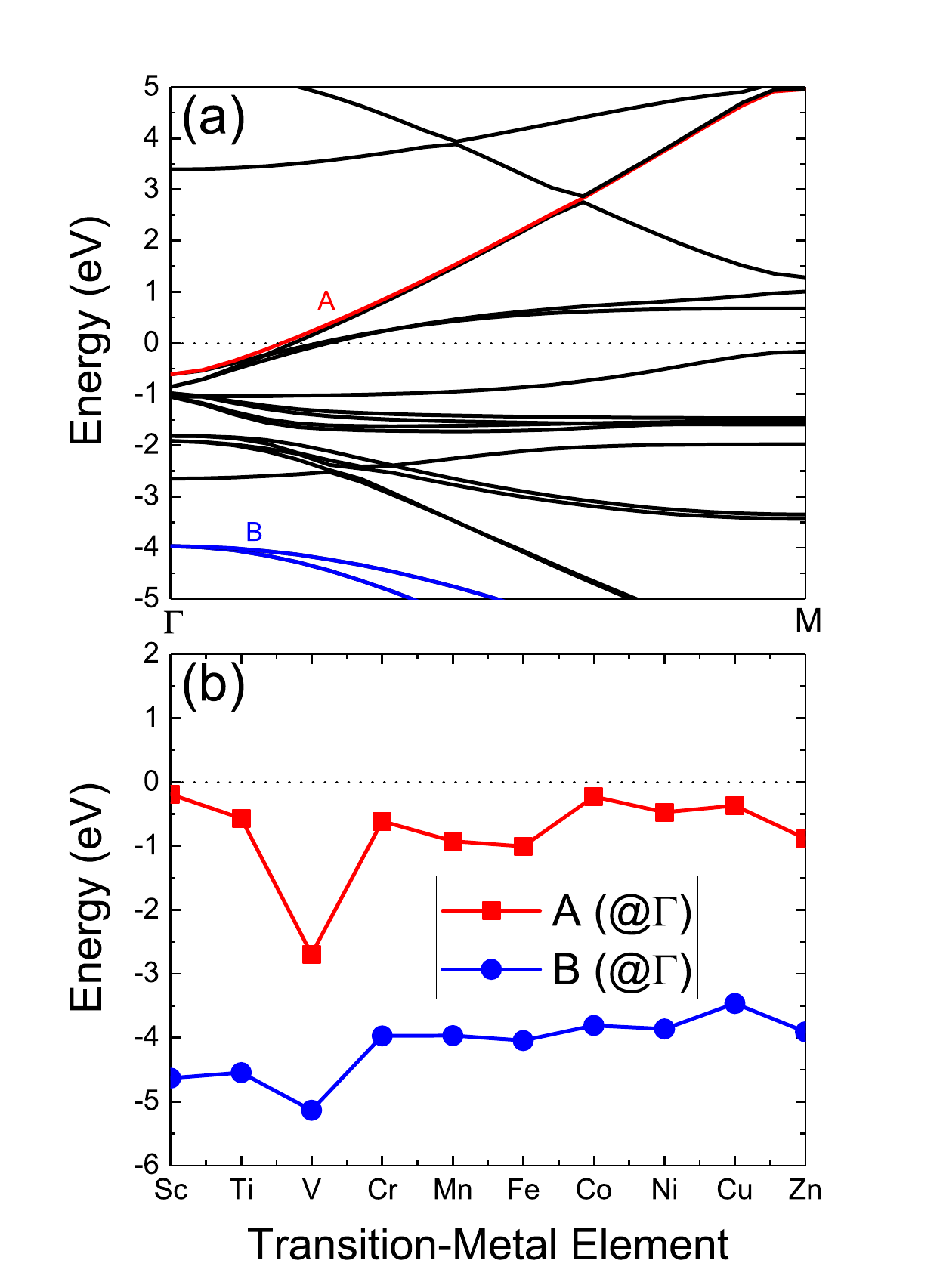}
\caption{According to the Cambridge conjecture, the main predictor for superconductivity used for other systems has been interlayer band crossing coming from $s$- or $p$-bands. This is observed in the transitions -metal cases as well. However, for our current case, a) also depicts the presence of $d$-band crossing between the $k$-pathway from $\Gamma-M$, which could indicate $spd$ hybridization. b) Energy shift of A and B bands from the Fermi energy at the $\Gamma-$point.}
\label{A-B}
\end{figure}

\begin{figure}
\includegraphics[width=1.0\linewidth]{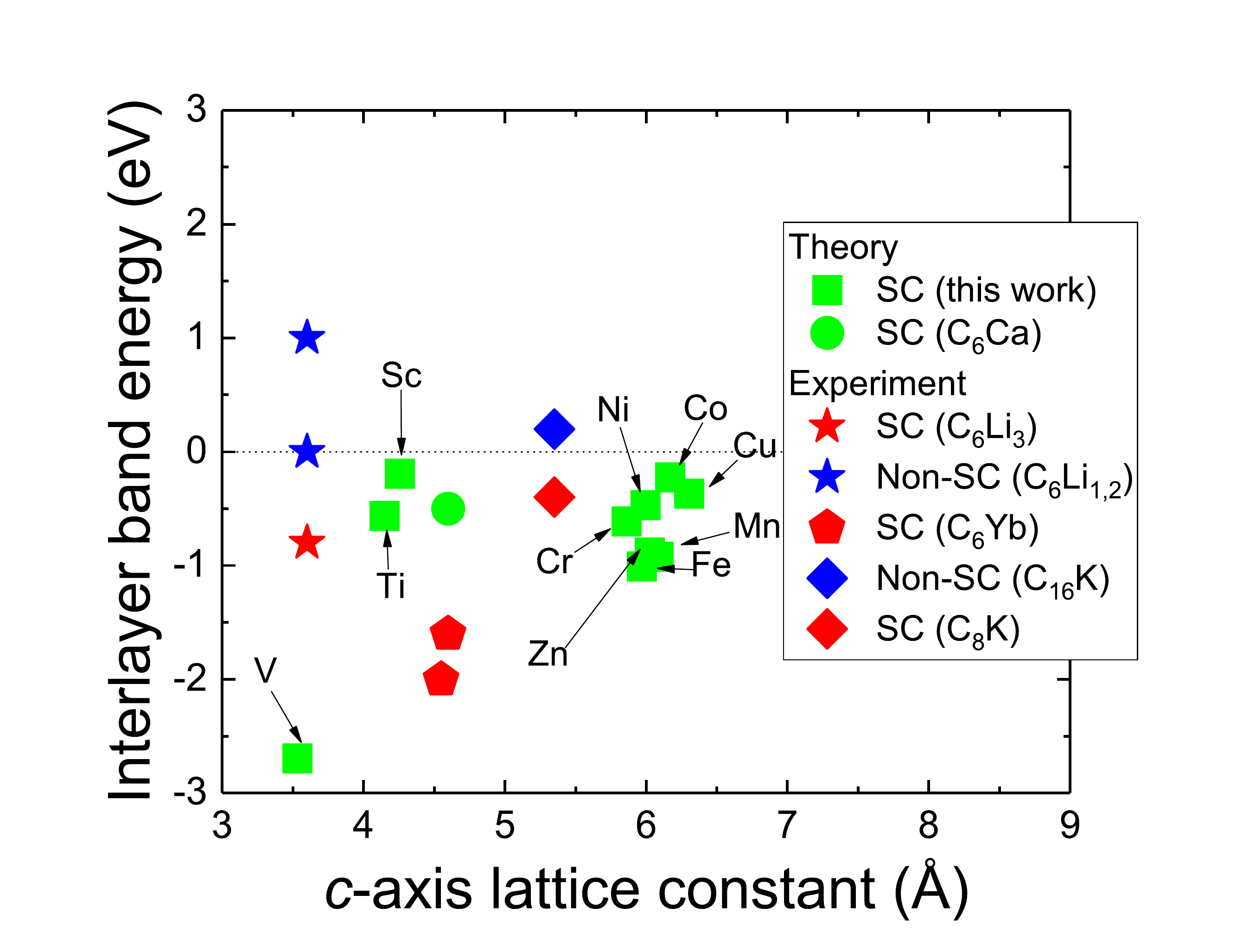}
\caption{A comparison of the interlayer band energy versus the c-axis lattice constant. Listed are the transition metal structures in this study and the experimental and theoretical structures for Li, K, Ca, and Yb\cite{Weller:05,Csanyi:05,mazin:11,Calandra:05,Margine:2016,Kaneko:17,Lee:12,Tiwari:17,Ichinokura:16}.}
\label{Elattice}
\end{figure}

In our investigation of the magnetic properties of transition-metal intercalated bilayer graphene, we find a similar phenomenon in the electronic structure of transition metals compared to the superconducting alkali and alkaline earth metals. However, rather than the charge density contributing mainly to a $p$ band crossing, the intercalated transition-metal bilayer graphene displays a hybridization with $d$ bands crossing near the Fermi energy. For all first-row transition metals, we have found interlayer $d$ bands crossing the Fermi energy as shown in Figure \ref{A-B}, suggestive of superconductivity. In comparison with previously investigated systems, Fig. \ref{Elattice} shows the interlayer band energy as a function of the $c$-axis lattice constant, where the superconducting systems can be distinguished by having an interlayer band energy lower than zero. Here, all transition-metal system have this band whose energy is below the Fermi level.  V has the lowest lattice constant of 3.5 \AA, followed by Ti and Sc which both share a distance of about 4 \AA. The remaining transition metals, which all lie approximately 2 \AA$\;$higher in lattice spacing than V, Ti, and Sc, are clustered around 6 \AA. Of the known experimental and theoretical studies of bilayer intercalated graphene, the separation distance for V is most comparable to C$_6$Li$_3$ around 3.5 \AA$\;$and Ti and Si are similar to C$_6$Yb and C$_6$Ca approximately 4.5 \AA. For the remaining transition metals, their lattice spacing is closer to the distance of C$_6$K which lies around 5.5 \AA.

Although these separation distances are quite similar to the superconducting metals, the main question is what effect does the intrinsic magnetism introduce into the system. As shown in Figure \ref{ED-MM-DE}, ferromagnetism and antiferromagnetism vary significantly across all TM atoms, indicating that their magnetic state does not seem to affect the overall levels. Instead, the main factor appears to be the hybridization of orbitals between the graphene sheets and the transition metals which lowers the $d$ orbital bands into the range of the Fermi energy. For Sc, Ti, and V, their $d$ orbital electrons strong hybridization with the graphene layers is enough to lower the interlayer band energy below the Fermi energy. This effect can be seen by the fact that the Dirac cone of the graphene layers is broken and falls to 2.5 eV below the Fermi energy for Sc and Ti and 3.5 eV for V as shown in Figure \ref{Sc-Mn-BS-DOS}. 

For the remaining transition metals, increasing the amount of $d$ orbital electrons diminishes the hybridization of the $d$ orbitals with the resonant $p$ orbitals of the graphene layers, hence expanding the layer separation. This effect is apparent as the Dirac cone appears to reemerge and shift up closer to the Fermi energy which has been demonstrated before for physisorbed graphene sheets on metal surfaces.\cite{Khom:09} Beyond V, the TM elements Cr, Fe, and Mn have the most robust hybridization indicated by Figure \ref{A-B} with some of the lowest band crossing energies and by the distortion of the Dirac cone in  Figure \ref{Sc-Mn-BS-DOS} and \ref{Fe-Zn-BS-DOS}. For Co, Ni, Cu, and Zn, the physical interactions dominate as the Dirac cone begins to reform, and appears fully formed for Zn and Cu. However, these physical interactions are enough to maintain the $d$ orbital band crossing at the Fermi energy, and surprisingly even lower the bands as seen in Zn.

\section{Conclusions}

Using first-principle methods and incorporating a nearest neighbor spin interaction picture, we studied the magnetic properties of intercalated bilayer graphene and the effects on the electronic properties with all $3d$ TMs to predict Dirac materials and superconductivity. Analyzing the spin-polarized electronic distributions, V and Ni preferred a ferromagnetic arrangement of spins while Cu and Zn had no magnetic moment present; the remaining TMs all had an antiferromagnetic arrangement of spins. We find that the electronic properties of the intercalated bilayer graphene lower the energy of the electronic Dirac cone and produces an electron pocket and consisting of possible hybridized carbon $p$ and TM $d$ bands. These interactions can be distinguished based upon the modifications of the electronic band structure and the layer separation. 

More importantly, we find that the characteristics of the $d$-bands below the Fermi level from the TMs are similar to those observed and predicted in intercalated alkaline and alkali metals in graphite. Therefore, it is possible that intercalation of the bilayer graphene with TM atoms may lead to a superconducting state. If shown experimentally, this may help in understanding the onset of superconductivity in graphene as well as shed light on the overall mechanisms and interactions that lead to such an interesting state.

\section*{Acknowledgements}
J.L.M-C.~gratefully acknowledges the support from the Energy and Materials Initiative at FSU. J.L.M.-C. and K.L. thank the High Performance Computer cluster (HPC) at the Research Computing Center (RCC) in FSU for providing computational resources and support. A portion of this work was performed at the National High Magnetic Field Laboratory (NHMFL), which is supported by National Science Foundation Cooperative Agreement No. DMR-1644779 and the State of Florida. A.M., A.A., and J.T.H acknowledge support from the Institute for Materials Science at Los Alamos National Laboratory.

\clearpage
\newpage
\onecolumngrid
\linespread{1.1}

\begin{center}
	{\Huge Supplementary Information}

\vspace{3pt}
\hspace{-22pt} \rule{1.0\textwidth}{2.5pt}

\Large{\textbf{Designing a path towards superconductivity through magnetic exchange in transition-metal intercalated bilayer graphene}}

\Large
Kevin P. Lucht,$^{1, 2, 3}$ A. D. Mahabir,$^{4}$ A. Alcantara,$^{4}$ A. V. Balatsky,$^{5}$
Jose L. Mendoza-Cortes,$^{1, 2, 3}$ and J. T. Haraldsen$^{4}$
\vspace{20pt}

\begin{flushleft}
	\normalsize
	
	$^{1}$Condensed Matter Theory - National High Magnetic Field Laboratory (NHMFL), Florida State University, Tallahassee, FL, 32310, USA\\
	
	$^{2}$Department of Scientific Computing, Materials Science and Engineering, High Performance Materials Institute (HPMI), Florida State University, Tallahassee, Florida, 32310, USA.\\
	 
	$^{3}$Department of Physics, College of Arts and Science, Florida State University, Tallahassee FL, 32310, USA.\\
	
	$^{4}$Department of Physics, University of North Florida, Jacksonville, FL 32224, USA\\
	
	$^{5}$Nordic Institute for Theoretical Physics, KTH Royal Institute of Technology and Stockholm University, Roslagstullsbacken 23, 106 91 Stockholm, Sweden
\end{flushleft}

\vspace{3pt}
\hspace{-22pt} \rule{1.0\textwidth}{2.5pt}

\normalsize

E-mail: 
mendoza@magnet.fsu.edu,\\
j.t.haraldsen@unf.edu

\end{center}

\clearpage
\newpage
\tableofcontents
\addcontentsline{toc}{section}{Suplementary Information}

\pagestyle{plain}
\clearpage
\newpage
\section{Optimized Structures (QuantumATK format)}

\subsection{Sc-BLG-SGGA-AFM (Ground State)}
\begin{verbatim}
Total free energy = -2244.97344 eV
+----------------------------------------------------------+
| Bulk Bravais lattice |
+----------------------------------------------------------+
Type:
Hexagonal

Lattice constants:
a = 4.305060 Ang
b = 4.305060 Ang
c = 20.000000 Ang

Lattice angles:
alpha = 90.000000 deg
beta = 90.000000 deg
gamma = 120.000000 deg

Primitive vectors:
u_1 = 2.152530 -3.728291 0.000000 Ang
u_2 = 2.152530 3.728291 0.000000 Ang
u_3 = 0.000000 0.000000 20.000000 Ang

+----------------------------------------------------------+
| Bulk: Cartesian (Angstrom) / fractional |
+----------------------------------------------------------+
14
Bulk
C -1.436497e+00 2.483384e+00 2.128091e+00 -0.66672 -0.00063 0.10640
C -7.160316e-01 1.244910e+00 2.128092e+00 -0.33328 0.00063 0.10640
C 7.160357e-01 1.244910e+00 2.128109e+00 -0.00063 0.33328 0.10641
C 2.866388e+00 -1.491935e-06 2.131903e+00 0.66582 0.66582 0.10660
C 1.438673e+00 3.840077e-06 2.131903e+00 0.33418 0.33418 0.10660
C 1.436495e+00 2.483384e+00 2.128108e+00 0.00063 0.66672 0.10641
C 7.160357e-01 1.244910e+00 1.787189e+01 -0.00063 0.33328 0.89359
C -1.436497e+00 2.483384e+00 1.787191e+01 -0.66672 -0.00063 0.89360
C 2.866388e+00 -1.491902e-06 1.786810e+01 0.66582 0.66582 0.89340
C 1.438673e+00 3.840099e-06 1.786810e+01 0.33418 0.33418 0.89340
C -7.160316e-01 1.244910e+00 1.787191e+01 -0.33328 0.00063 0.89360
C 1.436495e+00 2.483384e+00 1.787189e+01 0.00063 0.66672 0.89359
Sc 2.152481e+00 1.186776e+00 6.300000e-09 0.34083 0.65915 0.00000
Sc 5.029225e-05 2.541517e+00 6.280000e-09 -0.34083 0.34085 0.00000
\end{verbatim}

\clearpage
\newpage
\subsection{Sc-BLG-SGGA-FM}
\begin{verbatim}
Total free energy = -2244.88893 eV

+----------------------------------------------------------+
| Bulk Bravais lattice |
+----------------------------------------------------------+
Type:
Hexagonal

Lattice constants:
a = 4.305060 Ang
b = 4.305060 Ang
c = 20.000000 Ang

Lattice angles:
alpha = 90.000000 deg
beta = 90.000000 deg
gamma = 120.000000 deg

Primitive vectors:
u_1 = 2.152530 -3.728291 0.000000 Ang
u_2 = 2.152530 3.728291 0.000000 Ang
u_3 = 0.000000 0.000000 20.000000 Ang

+----------------------------------------------------------+
| Bulk: Cartesian (Angstrom) / fractional |
+----------------------------------------------------------+
14
Bulk
C -1.435616e+00 2.482895e+00 2.126955e+00 -0.66645 -0.00049 0.10635
C -7.169057e-01 1.245405e+00 2.126960e+00 -0.33355 0.00049 0.10635
C 7.169168e-01 1.245428e+00 2.126949e+00 -0.00050 0.33355 0.10635
C 2.865252e+00 4.176115e-06 2.111874e+00 0.66555 0.66556 0.10559
C 1.439806e+00 -1.049279e-05 2.111861e+00 0.33445 0.33444 0.10559
C 1.435609e+00 2.482867e+00 2.126954e+00 0.00049 0.66645 0.10635
C 7.169169e-01 1.245428e+00 1.787305e+01 -0.00050 0.33355 0.89365
C -1.435616e+00 2.482895e+00 1.787304e+01 -0.66645 -0.00049 0.89365
C 2.865252e+00 4.184548e-06 1.788813e+01 0.66555 0.66556 0.89441
C 1.439806e+00 -1.049280e-05 1.788814e+01 0.33445 0.33444 0.89441
C -7.169058e-01 1.245405e+00 1.787304e+01 -0.33355 0.00049 0.89365
C 1.435609e+00 2.482867e+00 1.787305e+01 0.00049 0.66645 0.89365
Sc 2.152544e+00 1.274696e+00 -2.040000e-09 0.32905 0.67095 -0.00000
Sc -1.193985e-05 2.453600e+00 -6.340000e-09 -0.32905 0.32905 -0.00000
\end{verbatim}

\clearpage
\newpage
\subsection{Ti-BLG-SGGA-AFM (Ground State)}
\begin{verbatim}
Total free energy = -2241.34660 eV
+----------------------------------------------------------+
| Bulk Bravais lattice |
+----------------------------------------------------------+
Type:
Hexagonal

Lattice constants:
a = 4.305060 Ang
b = 4.305060 Ang
c = 20.000000 Ang

Lattice angles:
alpha = 90.000000 deg
beta = 90.000000 deg
gamma = 120.000000 deg

Primitive vectors:
u_1 = 2.152530 -3.728291 0.000000 Ang
u_2 = 2.152530 3.728291 0.000000 Ang
u_3 = 0.000000 0.000000 20.000000 Ang

+----------------------------------------------------------+
| Bulk: Cartesian (Angstrom) / fractional |
+----------------------------------------------------------+
14
Bulk
C -1.433378e+00 2.486284e+00 2.073553e+00 -0.66639 0.00048 0.10368
C -7.192812e-01 1.242022e+00 2.073597e+00 -0.33365 -0.00051 0.10368
C 7.191754e-01 1.242020e+00 2.073597e+00 0.00049 0.33362 0.10368
C 2.870356e+00 3.081982e-05 2.088807e+00 0.66674 0.66674 0.10444
C 1.434613e+00 2.956195e-05 2.088761e+00 0.33323 0.33324 0.10444
C 1.433270e+00 2.486282e+00 2.073585e+00 -0.00051 0.66636 0.10368
C 7.192746e-01 1.241959e+00 1.792639e+01 0.00052 0.33364 0.89632
C -1.433282e+00 2.486270e+00 1.792642e+01 -0.66636 0.00050 0.89632
C 2.870499e+00 -7.138559e-08 1.791119e+01 0.66677 0.66677 0.89556
C 1.434709e+00 -1.122376e-06 1.791124e+01 0.33326 0.33326 0.89556
C -7.191329e-01 1.241948e+00 1.792637e+01 -0.33360 -0.00049 0.89632
C 1.433426e+00 2.486279e+00 1.792644e+01 -0.00047 0.66640 0.89632
Ti 2.152487e+00 1.239949e+00 -1.605691e-04 0.33370 0.66628 -0.00001
Ti -5.112495e-05 2.488376e+00 2.010060e-04 -0.33373 0.33370 0.00001
\end{verbatim}

\clearpage
\newpage
\subsection{Ti-BLG-SGGA-FM}
\begin{verbatim}
Total free energy = -2241.14776 eV
+----------------------------------------------------------+
| Bulk Bravais lattice |
+----------------------------------------------------------+
Type:
Hexagonal

Lattice constants:
a = 4.305060 Ang
b = 4.305060 Ang
c = 20.000000 Ang

Lattice angles:
alpha = 90.000000 deg
beta = 90.000000 deg
gamma = 120.000000 deg

Primitive vectors:
u_1 = 2.152530 -3.728291 0.000000 Ang
u_2 = 2.152530 3.728291 0.000000 Ang
u_3 = 0.000000 0.000000 20.000000 Ang

+----------------------------------------------------------+
| Bulk: Cartesian (Angstrom) / fractional |
+----------------------------------------------------------+
14
Bulk
C -1.429538e+00 2.486199e+00 2.038656e+00 -0.66548 0.00136 0.10193
C -7.229916e-01 1.242084e+00 2.038652e+00 -0.33452 -0.00136 0.10193
C 7.229926e-01 1.242092e+00 2.038654e+00 0.00136 0.33452 0.10193
C 2.868775e+00 1.338188e-06 2.060952e+00 0.66637 0.66637 0.10305
C 1.436285e+00 -4.994009e-06 2.060953e+00 0.33363 0.33363 0.10305
C 1.429539e+00 2.486190e+00 2.038653e+00 -0.00136 0.66548 0.10193
C 7.229807e-01 1.242084e+00 1.796134e+01 0.00136 0.33451 0.89807
C -1.429548e+00 2.486198e+00 1.796136e+01 -0.66549 0.00136 0.89807
C 2.868769e+00 -3.419962e-08 1.793904e+01 0.66637 0.66637 0.89695
C 1.436293e+00 1.583510e-05 1.793904e+01 0.33363 0.33363 0.89695
C -7.229789e-01 1.242102e+00 1.796133e+01 -0.33451 -0.00136 0.89807
C 1.429553e+00 2.486215e+00 1.796136e+01 -0.00136 0.66549 0.89807
Ti 2.152534e+00 1.277975e+00 3.320040e-06 0.32861 0.67139 0.00000
Ti -3.128556e-06 2.450327e+00 5.905800e-06 -0.32861 0.32861 0.00000
\end{verbatim}

\clearpage
\newpage
\subsection{V-BLG-SGGA-FM (Ground State)}
\begin{verbatim}
Total free energy = -2650.61888 eV
+----------------------------------------------------------+
| Bulk Bravais lattice |
+----------------------------------------------------------+
Type:
Hexagonal

Lattice constants:
a = 4.305060 Ang
b = 4.305060 Ang
c = 20.000000 Ang

Lattice angles:
alpha = 90.000000 deg
beta = 90.000000 deg
gamma = 120.000000 deg

Primitive vectors:
u_1 = 2.152530 -3.728291 0.000000 Ang
u_2 = 2.152530 3.728291 0.000000 Ang
u_3 = 0.000000 0.000000 20.000000 Ang

+----------------------------------------------------------+
| Bulk: Cartesian (Angstrom) / fractional |
+----------------------------------------------------------+
14
Bulk
C -1.430890e+00 2.487260e+00 1.766294e+00 -0.66594 0.00119 0.08831
C -7.216372e-01 1.241035e+00 1.766295e+00 -0.33406 -0.00119 0.08831
C 7.216379e-01 1.241037e+00 1.766290e+00 0.00119 0.33406 0.08831
C 2.872816e+00 1.423074e-06 1.770919e+00 0.66731 0.66731 0.08855
C 1.432246e+00 -1.626407e-06 1.770918e+00 0.33269 0.33269 0.08855
C 1.430893e+00 2.487258e+00 1.766289e+00 -0.00119 0.66594 0.08831
C 7.216379e-01 1.241037e+00 1.823371e+01 0.00119 0.33406 0.91169
C -1.430890e+00 2.487260e+00 1.823371e+01 -0.66594 0.00119 0.91169
C 2.872816e+00 1.423100e-06 1.822908e+01 0.66731 0.66731 0.91145
C 1.432246e+00 -1.626419e-06 1.822908e+01 0.33269 0.33269 0.91145
C -7.216372e-01 1.241035e+00 1.823370e+01 -0.33406 -0.00119 0.91169
C 1.430893e+00 2.487258e+00 1.823371e+01 -0.00119 0.66594 0.91169
V 2.152529e+00 1.256556e+00 -4.616000e-08 0.33148 0.66852 -0.00000
V 2.961059e-06 2.471746e+00 -4.610000e-08 -0.33148 0.33149 -0.00000
\end{verbatim}

\clearpage
\newpage
\subsection{V-BLG-SGGA-AFM}
\begin{verbatim}
Total free energy = -2650.59285 eV
+----------------------------------------------------------+
| Bulk Bravais lattice |
+----------------------------------------------------------+
Type:
Hexagonal

Lattice constants:
a = 4.305060 Ang
b = 4.305060 Ang
c = 20.000000 Ang

Lattice angles:
alpha = 90.000000 deg
beta = 90.000000 deg
gamma = 120.000000 deg

Primitive vectors:
u_1 = 2.152530 -3.728291 0.000000 Ang
u_2 = 2.152530 3.728291 0.000000 Ang
u_3 = 0.000000 0.000000 20.000000 Ang

+----------------------------------------------------------+
| Bulk: Cartesian (Angstrom) / fractional |
+----------------------------------------------------------+
14
Bulk
C -1.427171e+00 2.491674e+00 1.842768e+00 -0.66567 0.00265 0.09214
C -7.253114e-01 1.236627e+00 1.842757e+00 -0.33432 -0.00264 0.09214
C 7.253424e-01 1.236639e+00 1.842750e+00 0.00264 0.33433 0.09214
C 2.875080e+00 -1.170654e-05 1.904965e+00 0.66784 0.66784 0.09525
C 1.430053e+00 4.264718e-07 1.904773e+00 0.33218 0.33218 0.09524
C 1.427205e+00 2.491666e+00 1.842738e+00 -0.00264 0.66567 0.09214
C 7.253562e-01 1.236578e+00 1.815728e+01 0.00265 0.33433 0.90786
C -1.427157e+00 2.491727e+00 1.815729e+01 -0.66567 0.00266 0.90786
C 2.875180e+00 2.096202e-06 1.809515e+01 0.66786 0.66786 0.90476
C 1.429830e+00 -1.719179e-05 1.809514e+01 0.33213 0.33213 0.90476
C -7.253950e-01 1.236564e+00 1.815720e+01 -0.33433 -0.00266 0.90786
C 1.427114e+00 2.491742e+00 1.815718e+01 -0.00267 0.66566 0.90786
V 2.152533e+00 1.237179e+00 -5.263660e-06 0.33408 0.66592 -0.00000
V 3.573632e-06 2.491109e+00 -3.323160e-06 -0.33408 0.33408 -0.00000
\end{verbatim}

\clearpage
\newpage
\subsection{Cr-BLG-SGGA-AFM (Ground State)}
\begin{verbatim}
Total free energy = -2412.06218 eV
+----------------------------------------------------------+
| Bulk Bravais lattice |
+----------------------------------------------------------+
Type:
Hexagonal

Lattice constants:
a = 4.305060 Ang
b = 4.305060 Ang
c = 20.000000 Ang

Lattice angles:
alpha = 90.000000 deg
beta = 90.000000 deg
gamma = 120.000000 deg

Primitive vectors:
u_1 = 2.152530 -3.728291 0.000000 Ang
u_2 = 2.152530 3.728291 0.000000 Ang
u_3 = 0.000000 0.000000 20.000000 Ang

+----------------------------------------------------------+
| Bulk: Cartesian (Angstrom) / fractional |
+----------------------------------------------------------+
14
Bulk
C -1.435585e+00 2.486490e+00 2.930691e+00 -0.66693 -0.00000 0.14653
C -7.169407e-01 1.241801e+00 2.930691e+00 -0.33307 0.00000 0.14653
C 7.169430e-01 1.241809e+00 2.930694e+00 -0.00000 0.33307 0.14653
C 2.871137e+00 3.756332e-06 2.930644e+00 0.66692 0.66692 0.14653
C 1.433928e+00 -5.429272e-06 2.930645e+00 0.33308 0.33308 0.14653
C 1.435591e+00 2.486483e+00 2.930692e+00 0.00000 0.66693 0.14653
C 7.169207e-01 1.241800e+00 1.706931e+01 -0.00001 0.33307 0.85347
C -1.435607e+00 2.486482e+00 1.706931e+01 -0.66693 -0.00001 0.85347
C 2.871114e+00 -4.059371e-06 1.706936e+01 0.66692 0.66692 0.85347
C 1.433951e+00 2.338478e-06 1.706935e+01 0.33308 0.33309 0.85347
C -7.169188e-01 1.241809e+00 1.706931e+01 -0.33307 0.00001 0.85347
C 1.435614e+00 2.486492e+00 1.706931e+01 0.00001 0.66693 0.85347
Cr 2.152458e+00 1.242584e+00 4.428000e-07 0.33334 0.66663 0.00000
Cr -6.957971e-05 2.485498e+00 -1.042980e-06 -0.33335 0.33331 -0.00000
\end{verbatim}

\clearpage
\newpage
\subsection{Cr-BLG-SGGA-FM}
\begin{verbatim}
Total free energy = -2411.55867 eV
+----------------------------------------------------------+
| Bulk Bravais lattice |
+----------------------------------------------------------+
Type:
Hexagonal

Lattice constants:
a = 4.305060 Ang
b = 4.305060 Ang
c = 20.000000 Ang

Lattice angles:
alpha = 90.000000 deg
beta = 90.000000 deg
gamma = 120.000000 deg

Primitive vectors:
u_1 = 2.152530 -3.728291 0.000000 Ang
u_2 = 2.152530 3.728291 0.000000 Ang
u_3 = 0.000000 0.000000 20.000000 Ang

+----------------------------------------------------------+
| Bulk: Cartesian (Angstrom) / fractional |
+----------------------------------------------------------+
14
Bulk
C -1.435310e+00 2.486026e+00 2.983707e+00 -0.66680 -0.00000 0.14919
C -7.172366e-01 1.242277e+00 2.983714e+00 -0.33320 -0.00000 0.14919
C 7.172168e-01 1.242338e+00 2.983857e+00 -0.00001 0.33321 0.14919
C 2.870616e+00 4.519431e-05 2.983756e+00 0.66679 0.66681 0.14919
C 1.434430e+00 -2.600323e-05 2.983758e+00 0.33320 0.33319 0.14919
C 1.435305e+00 2.485964e+00 2.983848e+00 0.00001 0.66679 0.14919
C 7.172168e-01 1.242338e+00 1.701614e+01 -0.00001 0.33321 0.85081
C -1.435310e+00 2.486026e+00 1.701629e+01 -0.66680 -0.00000 0.85081
C 2.870616e+00 4.587769e-05 1.701624e+01 0.66679 0.66681 0.85081
C 1.434431e+00 -2.666100e-05 1.701624e+01 0.33320 0.33319 0.85081
C -7.172368e-01 1.242277e+00 1.701629e+01 -0.33320 -0.00000 0.85081
C 1.435305e+00 2.485964e+00 1.701615e+01 0.00001 0.66679 0.85081
Cr 2.152583e+00 1.242108e+00 7.496000e-07 0.33343 0.66659 0.00000
Cr 1.018774e-04 2.485655e+00 7.539200e-07 -0.33333 0.33337 0.00000
\end{verbatim}

\clearpage
\newpage
\subsection{Mn-BLG-SGGA-AFM (Ground State)}
\begin{verbatim}
Total free energy = -3120.06019 eV
+----------------------------------------------------------+
| Bulk Bravais lattice |
+----------------------------------------------------------+
Type:
Hexagonal

Lattice constants:
a = 4.305060 Ang
b = 4.305060 Ang
c = 20.000000 Ang

Lattice angles:
alpha = 90.000000 deg
beta = 90.000000 deg
gamma = 120.000000 deg

Primitive vectors:
u_1 = 2.152530 -3.728291 0.000000 Ang
u_2 = 2.152530 3.728291 0.000000 Ang
u_3 = 0.000000 0.000000 20.000000 Ang

+----------------------------------------------------------+
| Bulk: Cartesian (Angstrom) / fractional |
+----------------------------------------------------------+
14
Bulk
C -1.435336e+00 2.486085e+00 3.042950e+00 -0.66681 0.00000 0.15215
C -7.172019e-01 1.242213e+00 3.042944e+00 -0.33319 -0.00000 0.15215
C 7.172169e-01 1.242148e+00 3.042953e+00 0.00001 0.33318 0.15215
C 2.870645e+00 7.685082e-06 3.042982e+00 0.66681 0.66681 0.15215
C 1.434285e+00 3.026647e-05 3.042974e+00 0.33316 0.33317 0.15215
C 1.435369e+00 2.486095e+00 3.042946e+00 0.00000 0.66682 0.15215
C 7.171787e-01 1.242118e+00 1.695705e+01 0.00001 0.33317 0.84785
C -1.435369e+00 2.486053e+00 1.695705e+01 -0.66682 -0.00001 0.84785
C 2.870629e+00 -3.478247e-05 1.695702e+01 0.66681 0.66680 0.84785
C 1.434317e+00 5.302598e-05 1.695703e+01 0.33316 0.33318 0.84785
C -7.171711e-01 1.242250e+00 1.695706e+01 -0.33319 0.00001 0.84785
C 1.435395e+00 2.486127e+00 1.695706e+01 0.00001 0.66683 0.84785
Mn 2.152565e+00 1.242888e+00 1.883218e-03 0.33332 0.66669 0.00009
Mn 1.692063e-05 2.485491e+00 -1.820554e-03 -0.33332 0.33333 -0.00009
\end{verbatim}

\clearpage
\newpage
\subsection{Mn-BLG-SGGA-FM}
\begin{verbatim}
Total free energy = -3118.43674 eV
+----------------------------------------------------------+
| Bulk Bravais lattice |
+----------------------------------------------------------+
Type:
Hexagonal

Lattice constants:
a = 4.305060 Ang
b = 4.305060 Ang
c = 20.000000 Ang

Lattice angles:
alpha = 90.000000 deg
beta = 90.000000 deg
gamma = 120.000000 deg

Primitive vectors:
u_1 = 2.152530 -3.728291 0.000000 Ang
u_2 = 2.152530 3.728291 0.000000 Ang
u_3 = 0.000000 0.000000 20.000000 Ang

+----------------------------------------------------------+
| Bulk: Cartesian (Angstrom) / fractional |
+----------------------------------------------------------+
14
Bulk
C -1.435664e+00 2.486715e+00 1.994546e+00 -0.66698 0.00001 0.09973
C -7.168650e-01 1.241579e+00 1.994546e+00 -0.33302 -0.00001 0.09973
C 7.168659e-01 1.241579e+00 1.994546e+00 0.00001 0.33302 0.09973
C 2.871480e+00 1.818090e-06 1.994465e+00 0.66700 0.66700 0.09972
C 1.433582e+00 1.196606e-06 1.994465e+00 0.33300 0.33300 0.09972
C 1.435665e+00 2.486714e+00 1.994545e+00 -0.00001 0.66698 0.09973
C 7.168659e-01 1.241579e+00 1.800545e+01 0.00001 0.33302 0.90027
C -1.435664e+00 2.486715e+00 1.800545e+01 -0.66698 0.00001 0.90027
C 2.871480e+00 1.744591e-06 1.800553e+01 0.66700 0.66700 0.90028
C 1.433582e+00 9.561017e-07 1.800553e+01 0.33300 0.33300 0.90028
C -7.168647e-01 1.241579e+00 1.800545e+01 -0.33302 -0.00001 0.90027
C 1.435665e+00 2.486714e+00 1.800545e+01 -0.00001 0.66698 0.90027
Mn 2.152531e+00 1.241846e+00 -3.732000e-08 0.33346 0.66654 -0.00000
Mn 6.272365e-07 2.486447e+00 -5.234000e-08 -0.33346 0.33346 -0.00000
\end{verbatim}

\clearpage
\newpage
\subsection{Fe-BLG-SGGA-AFM (Ground State)}
\begin{verbatim}
Total free energy = -3402.01473 eV
+----------------------------------------------------------+
| Bulk Bravais lattice |
+----------------------------------------------------------+
Type:
Hexagonal

Lattice constants:
a = 4.305060 Ang
b = 4.305060 Ang
c = 20.000000 Ang

Lattice angles:
alpha = 90.000000 deg
beta = 90.000000 deg
gamma = 120.000000 deg

Primitive vectors:
u_1 = 2.152530 -3.728291 0.000000 Ang
u_2 = 2.152530 3.728291 0.000000 Ang
u_3 = 0.000000 0.000000 20.000000 Ang

+----------------------------------------------------------+
| Bulk: Cartesian (Angstrom) / fractional |
+----------------------------------------------------------+
14
Bulk
C -1.435266e+00 2.486305e+00 2.984138e+00 -0.66683 0.00005 0.14921
C -7.171846e-01 1.242155e+00 2.984068e+00 -0.33318 -0.00001 0.14920
C 7.172847e-01 1.242231e+00 2.984077e+00 0.00002 0.33321 0.14920
C 2.870659e+00 1.124271e-04 2.984605e+00 0.66680 0.66683 0.14923
C 1.434485e+00 3.979452e-05 2.984643e+00 0.33320 0.33321 0.14923
C 1.435356e+00 2.486217e+00 2.984117e+00 -0.00001 0.66684 0.14921
C 7.172845e-01 1.242252e+00 1.701592e+01 0.00002 0.33321 0.85080
C -1.435258e+00 2.486326e+00 1.701587e+01 -0.66683 0.00005 0.85079
C 2.870663e+00 1.215878e-04 1.701540e+01 0.66680 0.66683 0.85077
C 1.434481e+00 4.547964e-05 1.701536e+01 0.33320 0.33321 0.85077
C -7.171842e-01 1.242172e+00 1.701593e+01 -0.33318 -0.00000 0.85080
C 1.435347e+00 2.486234e+00 1.701589e+01 -0.00002 0.66684 0.85079
Fe 2.153751e+00 1.243446e+00 2.236576e-04 0.33353 0.66704 0.00001
Fe 9.643247e-04 2.480758e+00 -1.903689e-04 -0.33247 0.33292 -0.00001
\end{verbatim}

\clearpage
\newpage
\subsection{Fe-BLG-SGGA-FM}
\begin{verbatim}
Total free energy = -3401.91558 eV
+----------------------------------------------------------+
| Bulk Bravais lattice |
+----------------------------------------------------------+
Type:
Hexagonal

Lattice constants:
a = 4.305060 Ang
b = 4.305060 Ang
c = 20.000000 Ang

Lattice angles:
alpha = 90.000000 deg
beta = 90.000000 deg
gamma = 120.000000 deg

Primitive vectors:
u_1 = 2.152530 -3.728291 0.000000 Ang
u_2 = 2.152530 3.728291 0.000000 Ang
u_3 = 0.000000 0.000000 20.000000 Ang

+----------------------------------------------------------+
| Bulk: Cartesian (Angstrom) / fractional |
+----------------------------------------------------------+
14
Bulk
C -1.435024e+00 2.485821e+00 3.084104e+00 -0.66671 0.00004 0.15421
C -7.174693e-01 1.242543e+00 3.084106e+00 -0.33329 -0.00002 0.15421
C 7.175109e-01 1.242546e+00 3.084081e+00 0.00003 0.33330 0.15420
C 2.870380e+00 3.585553e-05 3.085328e+00 0.66674 0.66675 0.15427
C 1.434725e+00 3.731110e-05 3.085328e+00 0.33326 0.33327 0.15427
C 1.435064e+00 2.485826e+00 3.084088e+00 -0.00003 0.66672 0.15420
C 7.175110e-01 1.242546e+00 1.691592e+01 0.00003 0.33330 0.84580
C -1.435024e+00 2.485822e+00 1.691590e+01 -0.66671 0.00004 0.84579
C 2.870381e+00 3.604747e-05 1.691467e+01 0.66674 0.66675 0.84573
C 1.434725e+00 3.762899e-05 1.691467e+01 0.33326 0.33327 0.84573
C -7.174694e-01 1.242544e+00 1.691589e+01 -0.33329 -0.00002 0.84579
C 1.435064e+00 2.485826e+00 1.691591e+01 -0.00003 0.66672 0.84580
Fe 2.151804e+00 1.222478e+00 1.002520e-06 0.33589 0.66378 0.00000
Fe -7.954283e-04 2.504037e+00 -1.071020e-06 -0.33600 0.33563 -0.00000
\end{verbatim}

\clearpage
\newpage
\subsection{Co-BLG-SGGA-AFM (Ground State)}
\begin{verbatim}
Total free energy = -3748.23767 eV
+----------------------------------------------------------+
| Bulk Bravais lattice |
+----------------------------------------------------------+
Type:
Hexagonal

Lattice constants:
a = 4.305060 Ang
b = 4.305060 Ang
c = 20.000000 Ang

Lattice angles:
alpha = 90.000000 deg
beta = 90.000000 deg
gamma = 120.000000 deg

Primitive vectors:
u_1 = 2.152530 -3.728291 0.000000 Ang
u_2 = 2.152530 3.728291 0.000000 Ang
u_3 = 0.000000 0.000000 20.000000 Ang

+----------------------------------------------------------+
| Bulk: Cartesian (Angstrom) / fractional |
+----------------------------------------------------------+
14
Bulk
C -1.435024e+00 2.485821e+00 3.084104e+00 -0.66671 0.00004 0.15421
C -7.174693e-01 1.242543e+00 3.084106e+00 -0.33329 -0.00002 0.15421
C 7.175109e-01 1.242546e+00 3.084081e+00 0.00003 0.33330 0.15420
C 2.870380e+00 3.585553e-05 3.085328e+00 0.66674 0.66675 0.15427
C 1.434725e+00 3.731110e-05 3.085328e+00 0.33326 0.33327 0.15427
C 1.435064e+00 2.485826e+00 3.084088e+00 -0.00003 0.66672 0.15420
C 7.175110e-01 1.242546e+00 1.691592e+01 0.00003 0.33330 0.84580
C -1.435024e+00 2.485822e+00 1.691590e+01 -0.66671 0.00004 0.84579
C 2.870381e+00 3.604747e-05 1.691467e+01 0.66674 0.66675 0.84573
C 1.434725e+00 3.762899e-05 1.691467e+01 0.33326 0.33327 0.84573
C -7.174694e-01 1.242544e+00 1.691589e+01 -0.33329 -0.00002 0.84579
C 1.435064e+00 2.485826e+00 1.691591e+01 -0.00003 0.66672 0.84580
Co 2.151804e+00 1.222478e+00 1.002520e-06 0.33589 0.66378 0.00000
Co -7.954283e-04 2.504037e+00 -1.071020e-06 -0.33600 0.33563 -0.00000
\end{verbatim}

\clearpage
\newpage
\subsection{Co-BLG-SGGA-FM}
\begin{verbatim}
Total free energy = -3746.34919 eV
+----------------------------------------------------------+
| Bulk Bravais lattice |
+----------------------------------------------------------+
Type:
Hexagonal

Lattice constants:
a = 4.305060 Ang
b = 4.305060 Ang
c = 20.000000 Ang

Lattice angles:
alpha = 90.000000 deg
beta = 90.000000 deg
gamma = 120.000000 deg

Primitive vectors:
u_1 = 2.152530 -3.728291 0.000000 Ang
u_2 = 2.152530 3.728291 0.000000 Ang
u_3 = 0.000000 0.000000 20.000000 Ang

+----------------------------------------------------------+
| Bulk: Cartesian (Angstrom) / fractional |
+----------------------------------------------------------+
14
Bulk
C -1.432543e+00 2.490418e+00 2.005472e+00 -0.66675 0.00123 0.10027
C -7.199647e-01 1.237864e+00 2.005519e+00 -0.33325 -0.00123 0.10028
C 7.199863e-01 1.237871e+00 2.005462e+00 0.00123 0.33325 0.10027
C 2.875895e+00 -1.179706e-07 2.009241e+00 0.66803 0.66803 0.10046
C 1.429216e+00 4.368715e-06 2.009117e+00 0.33198 0.33199 0.10046
C 1.432567e+00 2.490436e+00 2.005509e+00 -0.00123 0.66676 0.10028
C 7.199863e-01 1.237871e+00 1.799454e+01 0.00123 0.33325 0.89973
C -1.432543e+00 2.490418e+00 1.799453e+01 -0.66675 0.00123 0.89973
C 2.875895e+00 -6.436895e-08 1.799076e+01 0.66803 0.66803 0.89954
C 1.429216e+00 4.422302e-06 1.799088e+01 0.33198 0.33199 0.89954
C -7.199647e-01 1.237864e+00 1.799448e+01 -0.33325 -0.00123 0.89972
C 1.432567e+00 2.490436e+00 1.799449e+01 -0.00123 0.66676 0.89972
Co 2.152449e+00 1.263535e+00 1.413200e-07 0.33053 0.66943 0.00000
Co -1.333333e-04 2.464755e+00 -1.508000e-08 -0.33058 0.33052 -0.00000
\end{verbatim}

\clearpage
\newpage
\subsection{Ni-BLG-SGGA-FM (Ground State)}
\begin{verbatim}
Total free energy = -4145.71288 eV
+----------------------------------------------------------+
| Bulk Bravais lattice |
+----------------------------------------------------------+
Type:
Hexagonal

Lattice constants:
a = 4.305060 Ang
b = 4.305060 Ang
c = 20.000000 Ang

Lattice angles:
alpha = 90.000000 deg
beta = 90.000000 deg
gamma = 120.000000 deg

Primitive vectors:
u_1 = 2.152530 -3.728291 0.000000 Ang
u_2 = 2.152530 3.728291 0.000000 Ang
u_3 = 0.000000 0.000000 20.000000 Ang

+----------------------------------------------------------+
| Bulk: Cartesian (Angstrom) / fractional |
+----------------------------------------------------------+
14
Bulk
C -1.434963e+00 2.485739e+00 2.997897e+00 -0.66668 0.00004 0.14989
C -7.175634e-01 1.242545e+00 2.997897e+00 -0.33332 -0.00004 0.14989
C 7.175657e-01 1.242496e+00 2.997897e+00 0.00005 0.33331 0.14989
C 2.870641e+00 -2.869916e-05 2.997942e+00 0.66681 0.66680 0.14990
C 1.434421e+00 1.908999e-05 2.997942e+00 0.33319 0.33320 0.14990
C 1.434965e+00 2.485788e+00 2.997896e+00 -0.00005 0.66669 0.14989
C 7.175694e-01 1.242496e+00 1.700210e+01 0.00005 0.33331 0.85011
C -1.434959e+00 2.485739e+00 1.700210e+01 -0.66668 0.00004 0.85011
C 2.870644e+00 -2.879630e-05 1.700206e+01 0.66681 0.66680 0.85010
C 1.434418e+00 1.917462e-05 1.700206e+01 0.33319 0.33320 0.85010
C -7.175672e-01 1.242545e+00 1.700210e+01 -0.33332 -0.00004 0.85011
C 1.434962e+00 2.485788e+00 1.700210e+01 -0.00005 0.66669 0.85011
Ni 2.152624e+00 1.194867e+00 -1.798200e-07 0.33978 0.66027 -0.00000
Ni 9.790250e-05 2.533426e+00 -1.803000e-07 -0.33973 0.33978 -0.00000
\end{verbatim}

\clearpage
\newpage
\subsection{Ni-BLG-SGGA-AFM}
\begin{verbatim}
Total free energy = -4145.63645 eV
+----------------------------------------------------------+
| Bulk Bravais lattice |
+----------------------------------------------------------+
Type:
Hexagonal

Lattice constants:
a = 4.305060 Ang
b = 4.305060 Ang
c = 20.000000 Ang

Lattice angles:
alpha = 90.000000 deg
beta = 90.000000 deg
gamma = 120.000000 deg

Primitive vectors:
u_1 = 2.152530 -3.728291 0.000000 Ang
u_2 = 2.152530 3.728291 0.000000 Ang
u_3 = 0.000000 0.000000 20.000000 Ang

+----------------------------------------------------------+
| Bulk: Cartesian (Angstrom) / fractional |
+----------------------------------------------------------+
14
Bulk
C -1.434909e+00 2.485969e+00 3.003509e+00 -0.66670 0.00008 0.15018
C -7.176251e-01 1.242280e+00 3.003513e+00 -0.33330 -0.00009 0.15018
C 7.176201e-01 1.242297e+00 3.003517e+00 0.00009 0.33330 0.15018
C 2.870690e+00 2.283683e-06 3.004568e+00 0.66682 0.66682 0.15023
C 1.434344e+00 1.899144e-05 3.004570e+00 0.33317 0.33318 0.15023
C 1.434931e+00 2.486010e+00 3.003514e+00 -0.00009 0.66671 0.15018
C 7.176276e-01 1.242295e+00 1.699648e+01 0.00009 0.33330 0.84982
C -1.434903e+00 2.485975e+00 1.699649e+01 -0.66670 0.00009 0.84982
C 2.870679e+00 5.591893e-06 1.699543e+01 0.66681 0.66682 0.84977
C 1.434354e+00 1.607749e-05 1.699543e+01 0.33318 0.33318 0.84977
C -7.176330e-01 1.242274e+00 1.699649e+01 -0.33330 -0.00009 0.84982
C 1.434924e+00 2.486011e+00 1.699649e+01 -0.00009 0.66671 0.84982
Ni 2.152492e+00 1.210682e+00 1.163880e-06 0.33763 0.66236 0.00000
Ni -7.706333e-06 2.517615e+00 1.157360e-06 -0.33764 0.33763 0.00000
\end{verbatim}

\clearpage
\newpage
\subsection{Cu-BLG-SGGA-AFM (Ground State)}
\begin{verbatim}
Total free energy = -4348.98090 eV
+----------------------------------------------------------+
| Bulk Bravais lattice |
+----------------------------------------------------------+
Type:
Hexagonal

Lattice constants:
a = 4.305060 Ang
b = 4.305060 Ang
c = 20.000000 Ang

Lattice angles:
alpha = 90.000000 deg
beta = 90.000000 deg
gamma = 120.000000 deg

Primitive vectors:
u_1 = 2.152530 -3.728291 0.000000 Ang
u_2 = 2.152530 3.728291 0.000000 Ang
u_3 = 0.000000 0.000000 20.000000 Ang

+----------------------------------------------------------+
| Bulk: Cartesian (Angstrom) / fractional |
+----------------------------------------------------------+
14
Bulk
C -1.435028e+00 2.485524e+00 3.152250e+00 -0.66667 -0.00000 0.15761
C -7.175146e-01 1.242779e+00 3.152247e+00 -0.33334 0.00000 0.15761
C 7.175172e-01 1.242768e+00 3.152249e+00 0.00000 0.33334 0.15761
C 2.869980e+00 -9.227003e-06 3.151932e+00 0.66665 0.66665 0.15760
C 1.435077e+00 1.814260e-05 3.151931e+00 0.33334 0.33335 0.15760
C 1.435002e+00 2.485513e+00 3.152253e+00 -0.00000 0.66666 0.15761
C 7.175743e-01 1.242767e+00 1.684775e+01 0.00001 0.33335 0.84239
C -1.434971e+00 2.485525e+00 1.684775e+01 -0.66665 0.00001 0.84239
C 2.870041e+00 -9.569987e-06 1.684807e+01 0.66667 0.66667 0.84240
C 1.435017e+00 1.813691e-05 1.684807e+01 0.33333 0.33334 0.84240
C -7.175714e-01 1.242778e+00 1.684775e+01 -0.33335 -0.00001 0.84239
C 1.434945e+00 2.485514e+00 1.684775e+01 -0.00002 0.66665 0.84239
Cu 2.152543e+00 1.254685e+00 6.958000e-08 0.33174 0.66827 0.00000
Cu 1.417914e-05 2.473811e+00 4.636000e-08 -0.33176 0.33177 0.00000
\end{verbatim}

\clearpage
\newpage
\subsection{Cu-BLG-SGGA-FM}
\begin{verbatim}
Total free energy = -4348.98090 eV
+----------------------------------------------------------+
| Bulk Bravais lattice |
+----------------------------------------------------------+
Type:
Hexagonal

Lattice constants:
a = 4.305060 Ang
b = 4.305060 Ang
c = 20.000000 Ang

Lattice angles:
alpha = 90.000000 deg
beta = 90.000000 deg
gamma = 120.000000 deg

Primitive vectors:
u_1 = 2.152530 -3.728291 0.000000 Ang
u_2 = 2.152530 3.728291 0.000000 Ang
u_3 = 0.000000 0.000000 20.000000 Ang

+----------------------------------------------------------+
| Bulk: Cartesian (Angstrom) / fractional |
+----------------------------------------------------------+
14
Bulk
C -1.435028e+00 2.485524e+00 3.152250e+00 -0.66667 -0.00000 0.15761
C -7.175146e-01 1.242779e+00 3.152247e+00 -0.33334 0.00000 0.15761
C 7.175172e-01 1.242768e+00 3.152249e+00 0.00000 0.33334 0.15761
C 2.869980e+00 -9.227003e-06 3.151932e+00 0.66665 0.66665 0.15760
C 1.435077e+00 1.814260e-05 3.151931e+00 0.33334 0.33335 0.15760
C 1.435002e+00 2.485513e+00 3.152253e+00 -0.00000 0.66666 0.15761
C 7.175743e-01 1.242767e+00 1.684775e+01 0.00001 0.33335 0.84239
C -1.434971e+00 2.485525e+00 1.684775e+01 -0.66665 0.00001 0.84239
C 2.870041e+00 -9.569987e-06 1.684807e+01 0.66667 0.66667 0.84240
C 1.435017e+00 1.813691e-05 1.684807e+01 0.33333 0.33334 0.84240
C -7.175714e-01 1.242778e+00 1.684775e+01 -0.33335 -0.00001 0.84239
C 1.434945e+00 2.485514e+00 1.684775e+01 -0.00002 0.66665 0.84239
Cu 2.152543e+00 1.254685e+00 6.958000e-08 0.33174 0.66827 0.00000
Cu 1.417914e-05 2.473811e+00 4.636000e-08 -0.33176 0.33177 0.00000
\end{verbatim}

\clearpage
\newpage
\subsection{Zn-BLG-SGGA-AFM (Ground State)}
\begin{verbatim}
Total free energy = -4975.43173 eV
+----------------------------------------------------------+
| Bulk Bravais lattice |
+----------------------------------------------------------+
Type:
Hexagonal

Lattice constants:
a = 4.305060 Ang
b = 4.305060 Ang
c = 20.000000 Ang

Lattice angles:
alpha = 90.000000 deg
beta = 90.000000 deg
gamma = 120.000000 deg

Primitive vectors:
u_1 = 2.152530 -3.728291 0.000000 Ang
u_2 = 2.152530 3.728291 0.000000 Ang
u_3 = 0.000000 0.000000 20.000000 Ang

+----------------------------------------------------------+
| Bulk: Cartesian (Angstrom) / fractional |
+----------------------------------------------------------+
14
Bulk
C -1.435428e+00 2.486066e+00 3.012436e+00 -0.66683 -0.00002 0.15062
C -7.170987e-01 1.242225e+00 3.012437e+00 -0.33317 0.00002 0.15062
C 7.171018e-01 1.242228e+00 3.012437e+00 -0.00002 0.33317 0.15062
C 2.870662e+00 1.084485e-06 3.012147e+00 0.66681 0.66681 0.15061
C 1.434396e+00 1.484080e-06 3.012147e+00 0.33319 0.33319 0.15061
C 1.435431e+00 2.486069e+00 3.012436e+00 0.00002 0.66683 0.15062
C 7.171018e-01 1.242228e+00 1.698756e+01 -0.00002 0.33317 0.84938
C -1.435428e+00 2.486066e+00 1.698756e+01 -0.66683 -0.00002 0.84938
C 2.870662e+00 1.084482e-06 1.698785e+01 0.66681 0.66681 0.84939
C 1.434396e+00 1.484080e-06 1.698785e+01 0.33319 0.33319 0.84939
C -7.170987e-01 1.242225e+00 1.698756e+01 -0.33317 0.00002 0.84938
C 1.435431e+00 2.486069e+00 1.698756e+01 0.00002 0.66683 0.84938
Zn 2.152535e+00 1.243412e+00 5.800000e-10 0.33325 0.66675 0.00000
Zn 3.596049e-06 2.484880e+00 5.800000e-10 -0.33325 0.33325 0.00000
\end{verbatim}

\clearpage
\newpage
\subsection{Zn-BLG-SGGA-FM}

\begin{verbatim}
Total free energy = -4975.43173 eV
+----------------------------------------------------------+
| Bulk Bravais lattice |
+----------------------------------------------------------+
Type:
Hexagonal

Lattice constants:
a = 4.305060 Ang
b = 4.305060 Ang
c = 20.000000 Ang

Lattice angles:
alpha = 90.000000 deg
beta = 90.000000 deg
gamma = 120.000000 deg

Primitive vectors:
u_1 = 2.152530 -3.728291 0.000000 Ang
u_2 = 2.152530 3.728291 0.000000 Ang
u_3 = 0.000000 0.000000 20.000000 Ang

+----------------------------------------------------------+
| Bulk: Cartesian (Angstrom) / fractional |
+----------------------------------------------------------+
14
Bulk
C -1.435428e+00 2.486066e+00 3.012436e+00 -0.66683 -0.00002 0.15062
C -7.170987e-01 1.242225e+00 3.012437e+00 -0.33317 0.00002 0.15062
C 7.171018e-01 1.242228e+00 3.012437e+00 -0.00002 0.33317 0.15062
C 2.870662e+00 1.084485e-06 3.012147e+00 0.66681 0.66681 0.15061
C 1.434396e+00 1.484080e-06 3.012147e+00 0.33319 0.33319 0.15061
C 1.435431e+00 2.486069e+00 3.012436e+00 0.00002 0.66683 0.15062
C 7.171018e-01 1.242228e+00 1.698756e+01 -0.00002 0.33317 0.84938
C -1.435428e+00 2.486066e+00 1.698756e+01 -0.66683 -0.00002 0.84938
C 2.870662e+00 1.084482e-06 1.698785e+01 0.66681 0.66681 0.84939
C 1.434396e+00 1.484080e-06 1.698785e+01 0.33319 0.33319 0.84939
C -7.170987e-01 1.242225e+00 1.698756e+01 -0.33317 0.00002 0.84938
C 1.435431e+00 2.486069e+00 1.698756e+01 0.00002 0.66683 0.84938
Zn 2.152535e+00 1.243412e+00 5.800000e-10 0.33325 0.66675 0.00000
Zn 3.596049e-06 2.484880e+00 5.800000e-10 -0.33325 0.33325 0.00000
\end{verbatim}

\end{document}